\documentclass[fleqn,twoside]{article}
\usepackage{espcrc2}

\usepackage{graphicx}
\usepackage[figuresright]{rotating}

\newcommand{\AmS}{{\protect\the\textfont2  A\kern-.1667em\lower.5ex\hbox{M}\kern-.125emS}}

\title{Measurement of the Strong Coupling \as\ from the 
Four-Jet Rate in \epem\ annihilation}

\author{Jochen Schieck \address{Max-Planck-Institut f\"ur Physik\\ F\"ohringer Ring 6, 80805 M\"unchen, Germany} }

\def\Journal#1#2#3#4{{#1} {\bf #2} (#3) #4}

\def\NIM{\em Nucl. Inst. Meth.}
\def\NIMA{{\em Nucl. Inst. Meth.} A}

\def\PLB{{\em Phys. Lett.}  B}

\def\PRD{{\em Phys. Rev.} D}

\def\CPC{\em Comp. Phys. Comm.}

\def\etal{{\it et al.}}

\def\NPBPS{\em Nucl.\ Phys.\ B (Proc.\ Suppl.)}

\def\be{\begin{equation}}
\def\ee{\end{equation}}

\newcommand{\as}                {\ensuremath{\alpha_\mathrm{S}}}
\newcommand{\asmz}   {\ensuremath{\alpha_s(M_\mathrm{Z})}}
\newcommand{\epem}              {\ensuremath{\mathrm{e^+e^-}}}
\newcommand{\rs}                {\ensuremath{\sqrt{s}}}

\newcommand{\py}                {PYTHIA}
\newcommand{\hw}                {HERWIG}
\newcommand{\ar}                {ARIADNE}
\newcommand{\jt}                {JETSET}

\newcommand{\kw}                {KORALW}
\newcommand{\grc}  		{GRC4f}

\newcommand{\ycut}              {\ensuremath{y_{\mathrm{cut}}}}
\newcommand{\result} {\ensuremath{\asmz=0.1208\pm0.0006(stat.)\pm0.0021(exp.)\pm0.0019(had.)\pm0.0024(scale)}}
\newcommand{\restot} {\ensuremath{\asmz=0.1208\pm0.0038~(\mathrm{total~error})}}

\newcommand{\invpb}     {\ensuremath{\mathrm{pb}^{-1}}}
\newcommand{\zzero}     {\ensuremath{\mathrm{Z^0}}}

\newcommand{\mz}                {\ensuremath{M_{\mathrm{Z^0}}}}
\newcommand{\ww}                {\ensuremath{\mathrm{W^+W^-}}}

\newcommand{\ascu}    {\ensuremath{\alpha_\mathrm{S}^{\mathrm{3}}}}

\begin{document}

\begin{abstract}
Data from \epem\ annihilation into hadrons at centre-of-mass energies 
between 91~GeV and 209~GeV are used to study the four-jet rate as 
a function of the Durham algorithm's resolution parameter \ycut. The four-jet
rate is compared to next-to-leading order calculations that include the 
resummation of large logarithms. The strong coupling measured from the 
four-jet rate is 
\[
\result
\]
in agreement with the world average.
\vspace{1pc}
\end{abstract}

\maketitle

\section{Introduction}
The annihilation of an electron and a positron into hadrons allows
a precise test of Quantum Chromodynamics (QCD).
Multijet rates are predicted in perturbation theory as functions
of the jet-resolution parameter, with one free parameter, the
strong coupling \as.
Calculations beyond leading order are made possible by theoretical
developments achieved in the last few years. 
For multi-jet rates as well as numerous event shape 
distributions with perturbative expansions
starting at $\cal{O}(\as)$, matched next-to-leading order (NLO) and next-to-leading
logarithmic approximations (NLLA) provide very precise description of the
data over a wide range of the available kinematic region and centre-of-mass
energy~\cite{CTWT,CDOTW,DSjets,NTqcd98}. \\
In this analysis we use data collected by the OPAL Collaboration at LEP
at 13 centre-of-mass energy points covering the full LEP energy range of
91-209~GeV.
A more detailed description of the analysis can be found in~\cite{OPALPN}.
\section{Observable}
\label{theory}
Jet algorithms are applied to cluster the large number
of particles of an hadronic event into a small number of jets,
reflecting the parton structure of the event. For this
analysis we use the Durham scheme~\cite{CDOTW}. 
The next-to-leading order calculation predicts the four-jet 
rate $R_{4}$, which is the fraction of four-jet events, as a function
of \as~\cite{zoltan}.
The fixed-order perturbative
prediction is not reliable for small values of $y_{\mathrm{cut}}$.
An all-order resummation, given in Ref.~\cite{CDOTW}, is possible for
the Durham clustering algorithm.
This NLLA calculation is combined with the NLO-prediction using the 
modified ``R matching'' scheme described in Ref.~\cite{zoltan}.
\section{Analysis Procedure}
The data used in this analysis were collected by OPAL~\cite{OPALtech}
between 1995 and 2000 and correspond to 14.5 \invpb\ of 91 GeV data, 
10.4 \invpb\ of LEP1.5 data, with
centre-of-mass energies $\sqrt{s}$=130 GeV and 136 GeV and 
706.4 \invpb\ of LEP2 data with
$\sqrt{s}$=161 to 209~GeV. The highest energy points have a 
spread of 1-2~GeV and are grouped into the main energy points.
The data at 91~GeV were taken during calibration runs at the \zzero\ peak at 
the beginning of each year and during data taking in 1996-2000.
The breakdown of the data samples, mean $\sqrt{s}$ and 
collected luminosity are
given in Table~\ref{lumi}. 
To study fewer energy points with improved statistical power, 
we combine the data above 91 GeV into three energy points by taking luminosity 
times hadronic cross-section weighted averages of the samples.
The LEP1.5 data samples provide an energy point at 133~GeV, while the LEP2 samples
give energy points at 178~GeV and 198~GeV corresponding to the energy ranges from
161 to 183~GeV and from 189 to 209~GeV.\\
Samples of Monte Carlo simulated events (\jt\ 7.4\cite{jetset},  KK2f\cite{kk2f},
\py\ 6.125\cite{pythia}, \hw\ 6.2\cite{herwig}, \kw\ 1.42\cite{koralw} and 
\grc\cite{grc4f}) were used to correct the data for experimental resolution, acceptance and backgrounds. 
The Monte Carlo samples generated at each energy point studied were processed through 
a full simulation of the OPAL detector~\cite{gopal}, and reconstructed in the same
way as the data.
In addition, for comparisons with the corrected data, and when 
correcting for the effects of fragmentation, large samples
of generator-level Monte-Carlo events were employed, using
the parton shower models \py\ 6.158, \hw\ 6.2 and \ar\ 4.11 \cite{ariadne}. 
\subsection{Selection of Events}
The selection of events for this analysis consists of three main stages: the
identification of hadronic event candidates, the removal of events with a 
large amount of initial state radiation (ISR), and the removal of four-fermion
background events (at energies above the \ww\ production threshold 
electroweak four-fermion production becomes a substantial background). \\
The number of selected candidate events obtained
after applying the selection cuts are given in Table~\ref{lumi}. 
After all cuts, the acceptance for
non-radiative signal events
ranged from $88.5\%$ at 91~GeV to $76.5\%$ at 207~GeV.
The residual four-fermion 
background was negligible below 161~GeV, and increased from $2.1\%$ at 
161~GeV to $6.2\%$ at 207~GeV.
\subsection{Corrections to the data}
\label{detectorcorrection}
The expected number of residual four-fermion background events, was
subtracted from the number of data events.
The effects of detector resolution, acceptance 
and of residual ISR were then accounted for by a 
bin-by-bin correction procedure.
The Monte Carlo ratio
of the hadron level to the detector level 
was used as a correction factor for the data. 
\begin{table}[ht]
\begin{center}
\begin{tabular}{|r|r|r|r|} \hline
average&  luminosity  & number of  & number of  \\
energy & (\invpb) & selected & events  \\
in GeV &&events & predicted by \\
&&& Monte Carlo \\
\hline
91.3  & 14.7 & 397452 & 396560.0  \\
\hline
130.1 &5.31 & 318 & 368.4\\
136.1 &5.95 & 312 & 329.7\\
\hline
161.3 &10.1 & 281 & 275.3 \\
172.1 &10.4 & 218 & 232.2\\
183.7 &57.7 & 1077 & 1083.5\\
\hline
188.6 &185.2 & 3086 & 3130.1\\
191.6 &29.5 & 514 & 473.0\\
195.5 &76.7 & 1137 & 1161.3\\
199.5 &79.3 & 1090 & 1030.8 \\
201.6 &38.0 & 519 & 526.5\\
204.9 &82.0 & 1130 & 1089.6\\
206.6 &137.1 & 1717 & 1804.1\\
\hline
\end{tabular}
\end{center}
\caption{
The average center-of-mass energy and integrated luminosity for each data
sample, together with the numbers of selected data
and expected Monte Carlo events. The horizontal lines indicate the four energy intervals.}
\label{lumi}
\end{table}
\section{Systematic Uncertainties}
\label{systematic}
Several sources of possible systematic uncertainties are studied. 
We investigate experimental uncertainties, effects due to
hadronisation and the theoretical uncertainty, associated with missing 
higher order terms in the theoretical prediction.
\section{Results}
\subsection{Four-Jet Rate Distributions}
\begin{figure}[ht]
\includegraphics[scale=0.38]{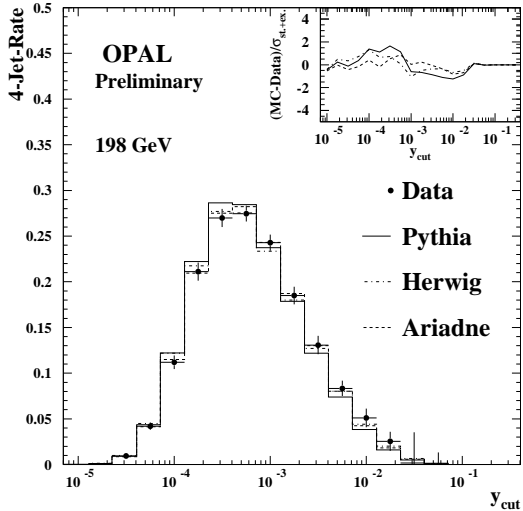}
\caption{The figures show the four-jet rate distribution
at hadron-level as a function of the $y_{\mathrm{cut}}$ resolution
parameter obtained with the Durham algorithm.}
\label{haddist} 
\end{figure}
The four-jet rates at $\sqrt{s}$=198~GeV after subtraction 
of background and correction for detector effects is shown 
in Figure~\ref{haddist}. 
Superimposed we show the predictions by the \py, \hw\ and \ar\ Monte
Carlo models, which in all cases were tuned to OPAL data at 91 GeV
centre-of-mass energy. In order to make a more sensitive comparison
between data and models, the inserts in the upper right corner show
the differences between data and each model, divided by the 
combined statistical and experimental error in that bin. 
The three models are seen to describe the high energy data at $\sqrt{s}$= 136, 178 and 198~GeV
well. 
However, some minor discrepancies are seen in the data taken at 91~GeV.
\subsection{Determination of \boldmath{\as}}{
\label{fitprocedure}
Our measurement of the strong coupling \as\ is based on the 
fits of QCD predictions to the corrected four-jet rate distribution.
The
combined $\cal{O}(\ascu)$+NLLA calculation was used. 
The theoretical predictions of the four-jet rate 
provide distributions at the parton level.
In order to confront the theory with the hadron-level data,
it is necessary to correct for hadronization
effects. The four-jet rate was calculated at hadron and parton
level using \py\ and, as a cross-check, with the \hw\ and \ar\ models. The theoretical
prediction is then multiplied by the ratio of the hadron- and parton-level 
four-jet rates.\\
A $\chi^{2}$-value for each energy point is calculated.
A single event can contribute to several $y_{\mathrm{cut}}$-bins
in a four-jet rate distribution and for this reason the bins
are correlated and the complete covariance matrix was 
determined.
The $\chi^{2}$ value is minimized with respect to \as\ for each
energy point separately. The scale parameter 
is set to unity.
The systematic
error was determined as described in Section~\ref{systematic}.
Each distribution was re-fitted and the difference to the default \as\ value
was taken as a symmetric systematic error. 
\begin{table}[ht]
\begin{center}
\begin{tabular}{|r|r|r|r|r|r|r|r|} \hline
 $\sqrt{s}$ & \as & stat. & exp. & hadr. & scale \\ 
GeV & & & & & \\
\hline
 $91.3$  & $ 0.1191 $ & $ 0.0001 $  & $ 0.0010$ & $ 0.0023$ & $ 0.0030$   \\
 $134.0$ & $ 0.1085 $ & $ 0.0038 $  & $ 0.0039$ & $ 0.0027$ & $ 0.0035$   \\
 $177.5$ & $ 0.1094 $ & $ 0.0022 $  & $ 0.0027$ & $ 0.0013$ & $ 0.0016$   \\
 $197.2$ & $ 0.1100 $ & $ 0.0009 $  & $ 0.0025$ & $ 0.0011$ & $ 0.0017$   \\
\hline
\end{tabular}
\end{center}
\caption{The mean value of \as\ for each energy interval, the statistical and
experimental errors, and the error due to hadronization and scale uncertainties. 
}
\label{fitcombinedtab}
\end{table}
\begin{figure}[ht]
\includegraphics[scale=0.380]{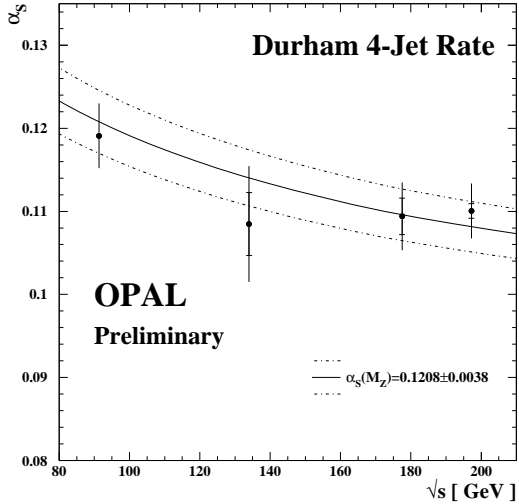}
\caption{The values for \as\ in the various energy intervals. The errors
show the statistical (inner part) and the total error. The statistical
error at 91~GeV is very small and cannot be seen.}
\label{alphas_run} 
\end{figure}
The result for \as(\rs) and the systematic uncertainties 
within each energy range are summarized 
in Table~\ref{fitcombinedtab}.
It is also of interest to combine the measurements of \as\
from the four different centre-of-mass energy points in
order to determine a single value.
This problem has been subject of extensive study by the LEP QCD working 
group~\cite{LEPQCDWG}, and we adopt their procedure here.\\
The set of \as\ measurements to be combined
are first evolved to a common scale, $Q=\mz$, assuming the validity of
QCD. The measurements are then combined in a weighted mean,
to minimize the $\chi^{2}$ between the combined values and the 
measurements. 
The result is 
$
\result, 
$
consistent with the world average value of $0.1172\pm0.0020$ ~\cite{alphaswa}.  
The average value of \as\ obtained and the 
results at each energy point are shown in Figure~\ref{alphas_run}.  
The hadronization and the scale uncertainty decrease with
increasing energy. The hadronization scales 
as an inverse power of $\sqrt{s}$ and the scale uncertainty 
varies with \ascu\ so that, for example, a $10\% $ smaller 
value of $\alpha_S$ at higher energies leads to a $27\% $ smaller scale 
uncertainty.  
\section{Summary}
In this note we present the preliminary measurements of the OPAL collaboration 
of the strong coupling from the four-jet rate 
at centre-of-mass energies between 91 and 209~GeV.
The predictions of the \py, \hw\ and \ar\ Monte Carlo models are found
to be in general agreement with the measured distributions. Some
differences are noted in the 91~GeV data where the statistics
are largest. \\
From a fit of $\cal{O}(\ascu)$+NLLA predictions to the four-jet rate,
we have determined the strong coupling \as. The value
of \asmz\ is determined to be \restot.

\end{document}